\documentclass[conference,10pt]{IEEEtran}
\IEEEoverridecommandlockouts
\usepackage{cite}
\usepackage{amsmath,amssymb,amsfonts}
\usepackage{algorithmic}
\usepackage{graphicx}
\usepackage{textcomp}
\usepackage{xcolor}
\usepackage{physics}
\usepackage{comment}
\usepackage{mathtools}
\usepackage{multirow}
\def\BibTeX{{\rm B\kern-.05em{\sc i\kern-.025em b}\kern-.08em
    T\kern-.1667em\lower.7ex\hbox{E}\kern-.125emX}}

\newcommand{\hermite}{\mathsf{H}}

\newcommand{\ceq}{\stackrel{c}{=}}

\newcommand{\Rdiag}{\vb{R}^\mathrm{diag}}
\newcommand{\dnndiag}{\mathrm{DNN}^\mathrm{diag}}
\newcommand{\dnntd}{\mathrm{DNN}^\text{time-domain}}

\newcommand{\Runc}{\vb{R}^\mathrm{rank1}}

\allowdisplaybreaks[3]

\begin{document}

\title{Independent Deeply Learned Tensor Analysis for Determined Audio Source Separation\thanks{This work was supported by JSPS-CAS Joint Research Program, Grant number JPJSBP120197203, and JSPS KAKENHI Grant Numbers JP20K19818, 19K20306, 19H01116, and 17H06101.}}

\author{\IEEEauthorblockN{Naoki Narisawa$^*$, Rintaro Ikeshita$^\dag$, Norihiro Takamune$^*$, Daichi Kitamura$^\ddag$,\\ Tomohiko Nakamura$^*$, Hiroshi Saruwatari$^*$, Tomohiro Nakatani$^\dag$}
\IEEEauthorblockA{$^*$Graduate School of Information Science and Technology, The University of Tokyo, Tokyo, Japan\\
$^\dag$NTT Communication Science Laboratories, NTT Corporation, Kyoto, Japan\\
$^\ddag$National Institute of Technology, Kagawa College, Kagawa, Japan}
}

\maketitle

\begin{abstract}
We address the determined audio source separation problem in the time-frequency domain.
In independent deeply learned matrix analysis (IDLMA), it is assumed that the inter-frequency correlation of each source spectrum is zero, which is inappropriate for modeling nonstationary signals such as music signals.
To account for the correlation between frequencies, independent positive semidefinite tensor analysis has been proposed.
This unsupervised (blind) method, however, severely restrict the structure of frequency covariance matrices (FCMs) to reduce the number of model parameters.
As an extension of these conventional approaches, we here propose a supervised method that models FCMs using deep neural networks (DNNs).
It is difficult to directly infer FCMs using DNNs.
Therefore, we also propose a new FCM model represented as a convex combination of a diagonal FCM and a rank-1 FCM.
Our FCM model is flexible enough to not only consider inter-frequency correlation, but also capture the dynamics of time-varying FCMs of nonstationary signals.
We infer the proposed FCMs using two DNNs: DNN for power spectrum estimation and DNN for time-domain signal estimation.
An experimental result of separating music signals shows that the proposed method provides higher separation performance than IDLMA.
\end{abstract}

\begin{IEEEkeywords}
audio source separation, independent component analysis, deep neural networks, inter-frequency correlation
\end{IEEEkeywords}

\section{Introduction}
Multichannel audio source separation is a technique of estimating source signals from their observed mixture using a microphone array. 
In particular, a separation problem that does not use any prior knowledge of the mixing system and property of source signals is called blind source separation (BSS)\cite{Sawada:APSIPA_J_SIP2019}. 
Frequency-domain independent component analysis (FDICA)\cite{Smaragdis:nc1998,Saruwatari:TASLP2006,Ono:LVAICA2010} and its multivariate extension,  independent vector analysis (IVA)\cite{Kim:ICA2006,Hiroe:ICA2006,Kim:TASLP2007,Ono:WASPAA2011}, are common BSS methods for a determined or overdetermined situation, i.e., the number of microphones is equal to or greater than that of sources. These methods deal with BSS in the time-frequency domain using the short-time Fourier transform (STFT) and rely on the statistical independence between sources to estimate the demixing matrix.

To improve the separation performance of FDICA and IVA, independent low-rank matrix analysis (ILRMA)\cite{Kitamura:IEEE_ACM_J_ASLP2016,Kitamura:springer2018,Kitamura:EURASIP_ASP2018} has been proposed for modeling the source power spectra using nonnegative matrix factorization (NMF)\cite{Lee:Nature1999,Fevotte:neco2009}. 
However, this low-rank modeling of source spectra is not appropriate for real-world signals such as speech and music signals \cite{Makishima:IEEE_ACM_J_ASLP2019}. 
In addition, for the NMF model, it is assumed that the correlation between the frequency bins is zero. In other words, the time series of frequency covariance matrices (FCMs) for each source is constrained to be a diagonal matrix, which is not appropriate because many nonstationary signals are known to have inter-frequency correlations.

As a supervised extension of ILRMA as well as FDICA, independent deeply learned matrix analysis (IDLMA)\cite{Makishima:IEEE_ACM_J_ASLP2019,Makishima:SigPro2021} has been proposed. 
Instead of using NMF, in IDLMA, a pretrained deep neural network (DNN) is used to model a source power spectrum. If sufficient data are available for training DNNs, DNNs can accurately estimate the source power spectrum during the demixing phase. However, in the same manner as in ILRMA, in IDLMA, it is still assumed that each FCM is diagonal, which is inappropriate for modeling nonstationary signals such as speech and music signals.

To consider the inter-frequency correlation of the source spectra, a BSS method called independent positive semidefinite tensor analysis (IPSDTA)\cite{Ikeshita:EUSIPCO2018,Kondo:ICASSP2020} has been proposed as an extension of ILRMA. In IPSDTA, positive semidefinite tensor factorization (PSDTF)\cite{Yoshii:ICML2013,Yoshii:ICASSP2016,Lintkus:MSLP2017}, which is an extension of NMF to capture inter-frequency correlation, is incorporated as the source spectrum model. Here, PSDTF is a method that models the time series of FCMs for each source as a convex cone of several FCM bases (which are Hermitian positive semidefinite). However, this PSDTF model is too restrictive to capture the nonstationarity of FCMs.

To overcome the drawbacks of the above conventional methods, we here propose a new supervised source separation method that accounts for the inter-frequency correlation using a flexible and tractable model of FCMs. 
We call the proposed method {\it independent deeply learned tensor analysis} (IDLTA) because the time series of FCMs, which is a tensor, is represented by DNNs. 
It is difficult to directly infer FCMs using DNNs. 
Therefore, we also propose to restrict each FCM to be a convex combination of the following two simple FCMs: 
(i) One is a diagonal FCM that corresponds to the power spectrum and can be estimated using DNNs in the same manner as in IDLMA. 
(ii) The other is a rank-1 FCM whose first eigenvector can be viewed as the original source signal and can be estimated using DNNs such as time-domain audio separation network (TasNet)\cite{Luo:ICASSP2018} and multiresolution deep layered analysis (MRDLA)\cite{Nakamura:ICASSP2020,Kozuka:INTERNOISE2020}.
We show that the rank-1 FCM is an optimal FCM in some respect (Section \ref{sec:model_opt}) but causes the numerical instability of IDLTA optimization.
Our proposed diagonal plus rank-1 FCM can compensate for this issue while capturing the inter-frequency correlation. We show in a numerical experiment that IDLTA provides higher separation performance than IDLMA for a music separation task.

\section{Audio source separation problem}
Suppose $N$ source signals are observed by $M$ microphones. 
The source and observed signals in the STFT domain are denoted as
\begin{align}
    \vb*{s}_{ij} &\coloneqq (s_{ij1}, ..., s_{ijn}, ..., s_{ijN})^\top\in \mathbb{C}^{N}, \\
    \vb*{x}_{ij} &\coloneqq (x_{ij1}, ..., x_{ijm}, ..., x_{ijM})^\top\in \mathbb{C}^{M}, 
\end{align}
respectively, where $\cdot^\top$ denotes the transpose, and $i=1,\ldots,I$, $j=1,\ldots,J$, $n=1,\ldots,N$, and $m=1,\ldots,M$ are indices of the frequency bins, time frames, sources, and microphones, respectively. In this paper, we suppose that the window size in the STFT is sufficiently longer than the impulse responses between sources and microphones. In this case, the mixing system can be expressed as
\begin{align}
    \vb*{x}_{ij} = \vb{A}_{i}\vb*{s}_{ij}\in\mathbb{C}^{M},
\end{align}
where $\vb{A}_i\in\mathbb{C}^{M\times N}$ is a time-invariant mixing matrix at the $i$th frequency bin. With the assumption that $N = M$ and $\vb{A}_i$ is invertible, the estimation of the original source $\vb*{s}_{ij}$, denoted as $\vb*{y}_{ij} \coloneqq (y_{ij1},\ldots, y_{ijN})^\top$, can be obtained as
\begin{align}
    \vb*{y}_{ij} = \vb{W}_i\vb*{x}_{ij}\in\mathbb{C}^N, \label{eq:y=Wx}
\end{align}
where $\vb{W}_i\coloneqq\pqty{\vb*{w}_{i1},\ldots,\vb*{w}_{iN}}^\hermite\in\mathbb{C}^{N\times M}$ is a demixing matrix, and $\cdot^\hermite$ denotes Hermitian transpose. 

\section{Conventional methods}\label{sec:conv_methods}
We summarize conventional methods in terms of FCMs and discuss their drawbacks in Section \ref{sec:issue}.

In conventional methods such as ILRMA \cite{Kitamura:IEEE_ACM_J_ASLP2016}, IDLMA\cite{Makishima:IEEE_ACM_J_ASLP2019}, and IPSDTA\cite{Ikeshita:EUSIPCO2018}, it is assumed that the sources are mutually independent. On the basis of this assumption and \eqref{eq:y=Wx}, the likelihood of the observed signals is described as \cite{Sawada:APSIPA_J_SIP2019}
\begin{align}
    p\pqty{\Bqty{\vb*{x}_{ij}}_{i,j}} 
    &= \prod_{n}p\pqty{\vb{Y}_n}\cdot\prod_{i, j}\abs{\det\vb{W}_i}^2, \label{eq:obs_l} \\
    \vb{Y}_n&\coloneqq\pqty{\va*{y}_{1n},\ldots,\va*{y}_{Jn}}\in\mathbb{C}^{I\times J}, \\
    \va*{y}_{jn}&\coloneqq\pqty{y_{1jn},\ldots,y_{Ijn}}^\top\in\mathbb{C}^I,
\end{align}
where $\va*{y}_{jn}$ is a separated signal of source $n$ and time frame $j$. It is also assumed that the random variables $\va*{y}_{1n},\ldots,\va*{y}_{Jn}$ are mutually independent and follow a multivariate complex Gaussian distribution with the zero mean and covariance matrix $\vb{R}_{jn}$ as follows:
\begin{align}
    p\pqty{\vb{Y}_n} = \prod_j\frac{1}{\pi^I\det\vb{R}_{jn}}\exp\pqty{-\va*{y}_{jn}^\hermite\vb{R}_{jn}^{-1}\va*{y}_{jn}}. \label{eq:prop_y}
\end{align}
Here, $\vb{R}_{jn}$ denotes the FCM. Then, the negative log-likelihood of the observed signal is derived from \eqref{eq:obs_l} and \eqref{eq:prop_y} as
\begin{align}
    \mathcal{L}
    \ceq
    &\sum_{j,n} \pqty{\log\det\vb{R}_{jn} + \va*{y}_{jn}^\hermite \vb{R}_{jn}^{-1}\va*{y}_{jn}} \notag\\
    &- J\sum_i\log\abs{\det \vb{W}_i}^2, \label{eq:bss_ll}
\end{align}
where $\ceq$ denotes equality up to a constant. The parameters to be estimated are the demixing matrices $\{\vb{W}_i\}_i$ and FCMs $\{\vb{R}_{jn}\}_{j,n}$.

As we explain in the following subsections, conventional methods such as ILRMA \cite{Kitamura:IEEE_ACM_J_ASLP2016}, IDLMA\cite{Makishima:IEEE_ACM_J_ASLP2019}, and IPSDTA\cite{Ikeshita:EUSIPCO2018} only differ in the way FCMs are modeled.

\subsection{ILRMA and IDLMA}\label{sec:ilrma_idlma}
In ILRMA \cite{Kitamura:IEEE_ACM_J_ASLP2016} and IDLMA \cite{Makishima:IEEE_ACM_J_ASLP2019}, the time series of the FCMs of source $n$, i.e., $\vb{R}_{1n},\ldots,\vb{R}_{Jn}$, is constrained to be diagonal:
\begin{align}
    \vb{R}_{jn} 
    =\Rdiag_{jn} 
    \coloneqq \mathrm{diag}\Bqty{\sigma_{1jn}^2,\ldots,\sigma_{Ijn}^2},
\end{align}
where $\mathrm{diag}\Bqty{\lambda_1,\ldots,\lambda_I}$ denotes a diagonal matrix with $\lambda_i$ as its $i$th diagonal element. Furthermore, in IDLMA, $\Rdiag_{jn}$, which corresponds to the power spectrum of source $n$,
is estimated using a pretrained DNN:
\begin{align}
    \{\Rdiag_{jn}\}_j = \dnndiag\pqty{\vb{Y}_n}.
\end{align}
Note that the input of $\dnndiag$ is the separated signal $\vb{Y}_n$ in IDLMA. In ILRMA, on the other hand, $\Rdiag_{jn}$ is modeled using NMF (see, e.g., \cite{Kitamura:IEEE_ACM_J_ASLP2016}).

\subsection{IPSDTA}
In IPSDTA \cite{Ikeshita:EUSIPCO2018}, the FCMs of each source $n$ are modeled by PSDTF \cite{Yoshii:ICML2013}, which is an extension of NMF:
\begin{align}
    \vb{R}_{jn} 
    = \sum_{k=1}^{K_n} h_{kjn}\vb{U}_{kn} . \label{eq:psdtf}
\end{align}
Here,
$h_{kjn}\ge 0$ and $\vb{U}_{kn}\in\mathbb{C}^{I\times I}$ are an activation and a time-invariant basis (Hermitian positive semidefinite matrix) for source $n$,
respectively, and $K_n$ is the number of bases of PSDTF.
Since $\vb{R}_{jn}$ is nondiagonal, IPSDTA can consider the inter-frequency correlation between frequency bins.

\subsection{Drawbacks of ILRMA, IDLMA, and IPSDTA}\label{sec:issue}
In ILRMA and IDLMA, FCMs are assumed to be diagonal, implying that the inter-frequency correlation of source spectra is constrained to be zero.
However, this assumption is not appropriate for nonstationary signals such as speech and music signals, because they have the inter-frequency correlation.

Unlike ILRMA and IDLMA, on the other hand, IPSDTA can consider the inter-frequency correlation by modeling FCMs by PSDTF. However, since PSDTF represents time-varying FCMs by a convex cone of a few positive definite bases, it severely restricts model flexibility and cannot accurately capture the nonstationarity of FCMs.

\section{Proposed supervised method: IDLTA}\label{sec:model}
To overcome the problems described in Section \ref{sec:issue} and improve the separation performance of conventional methods, we propose a new supervised method called IDLTA, which represents the time series of FCMs using DNNs and optimizes $\vb{W}_i$ in \eqref{eq:bss_ll} blindly.
Here, we discuss in Section \ref{sec:model_opt} that it is not appropriate to infer $\vb{R}_{jn}$ directly using DNNs, and in Section \ref{sec:model_prop}, we explain the new proposed FCM model.

\subsection{Optimal FCM} \label{sec:model_opt}
Consider the estimation of the FCM for the source signal $\va*{s}_{jn}:=\pqty{s_{1jn},\ldots,s_{Ijn}}^\top\in\mathbb{C}^I$.
Suppose that $\va*{s}_{jn}$ follows a multivariate complex Gaussian distribution with the zero mean and covariance matrix $\vb{R}_{jn}$ as follows:
\begin{align}
    \log p\pqty{\va*{s}_{jn}\mid\vb{R}_{jn}} \ceq -\va*{s}_{jn}^\hermite\vb{R}_{jn}^{-1}\va*{s}_{jn}-\log\det \vb{R}_{jn}.\label{eq:gen_s}
\end{align}
In this case, the maximum likelihood estimation of $\vb{R}_{jn}$ is given by $\vb{R}_{jn}=\va*{s}_{jn}\va*{s}_{jn}^\hermite$.
This implies that an optimal FCM is of rank-1.

Training DNN to maximize \eqref{eq:gen_s} is numerically unstable in the vicinity of the optimal FCM since \eqref{eq:gen_s} diverges to infinity when $\vb{R}_{jn}=\va*{s}_{jn}\va*{s}_{jn}^\hermite$.
In addition, when $\vb{R}_{jn}$ inferred using DNN is almost a rank-1 matrix, the separated signal $\va*{y}_{jn}$ is estimated only in the direction close to the eigenvector for the largest eigenvalue of $\vb{R}_{jn}$.
This is because the second term in the objective function \eqref{eq:bss_ll} with $\va*{y}_{jn}$ parallel to the other eigenvectors of $\vb{R}_{jn}$ becomes quite large and $\vb{W}_i$ is not estimated properly.
In this sense, it is not appropriate to estimate the optimal FCM directly using DNNs. 

\subsection{Proposed FCM}\label{sec:model_prop}
In our proposed model,  each FCM in \eqref{eq:obs_l}--\eqref{eq:prop_y} is represented as a convex combination of a diagonal FCM and a rank-1 FCM as follows:
\begin{align}
    \vb{R}_{jn} = \pqty{1-\alpha}\Rdiag_{jn} + \alpha\Runc_{jn}, \label{eq:R_convex}
\end{align}
where $0 \le\alpha \le 1$ and
\begin{align}
    &\Rdiag_{jn} 
    \coloneqq \text{diag}\Bqty{d_{1jn}^2,\ldots,d_{Ijn}^2}, \label{eq:def_D}\\
    &\Runc_{jn} \coloneqq \va*{z}_{jn}\va*{z}_{jn}^\hermite,\\
    &\va*{z}_{jn}
    \coloneqq \pqty{z_{1jn}, \ldots, z_{Ijn}}^\top\in\mathbb{C}^{I}. \label{eq:def_v}
\end{align}
The first term of \eqref{eq:R_convex}, i.e., $\Rdiag_{jn}$, can be viewed as the source power spectrum in the same manner as in IDLMA (see Section \ref{sec:ilrma_idlma}).
On the other hand, as we explain in Section \ref{sec:model_opt},
the second term of \eqref{eq:R_convex}, i.e., $\Runc_{jn}$, is equivalent to the source signal.
In particular, the proposed FCM with $\alpha=1$ is an optimal FCM in some respect. Since this $\vb{R}_{jn}$ has non-zero off-diagonal elements, it can be used to consider the inter-frequency correlation to some extent.

In the proposed IDLTA, $\Rdiag_{jn}$ is inferred using a pretrained DNN for estimating the power spectrum in the same manner as in IDLMA:
\begin{align}
    \{\Rdiag_{jn}\}_j = \dnndiag(\vb{Y}_n).
\end{align}
On the other hand, $\va*{z}_{jn}$ is inferred using a pretrained DNN for estimating the separated signals in the time domain:
\begin{align}
    \{\va*{z}_{jn}\}_j = \mathrm{STFT}\pqty{\mathrm{DNN}^\text{time-domain}\pqty{\mathrm{ISTFT}\pqty{\vb{Y}_n}}}. \label{eq:dnn_rank1}
\end{align}
Here, ISTFT means the inverse STFT. For instance, TasNet \cite{Luo:ICASSP2018} and MRDLA \cite{Nakamura:ICASSP2020,Kozuka:INTERNOISE2020} can be used for $\mathrm{DNN}^\text{time-domain}$ to estimate $\va*{z}_{jn}$.
Note that the input of both DNNs is the separated signal $\vb{Y}_n$ in IDLTA.

\subsection{Related research}
To reduce the computational cost of PSDTF, the acceleration of PSDTF has been proposed by approximating each basis $\vb{U}_{kn}$ in \eqref{eq:psdtf} as the sum of a diagonal matrix and a low-rank matrix \cite{Lintkus:MSLP2017}.
Since this approach further restricts the  structure of FCMs estimated by PSDTF, it cannnot capture the nonstationarity of FCMs in the same manner as in PSDTF.
In contrast, the proposed model of FCMs not only takes into account the time variability of FCMs but can also be estimated accurately using two DNNs.

\section{Algorithm for IDLTA}\label{sec:algo}
\begin{figure}
    \centering
    \includegraphics[width=.65\linewidth]{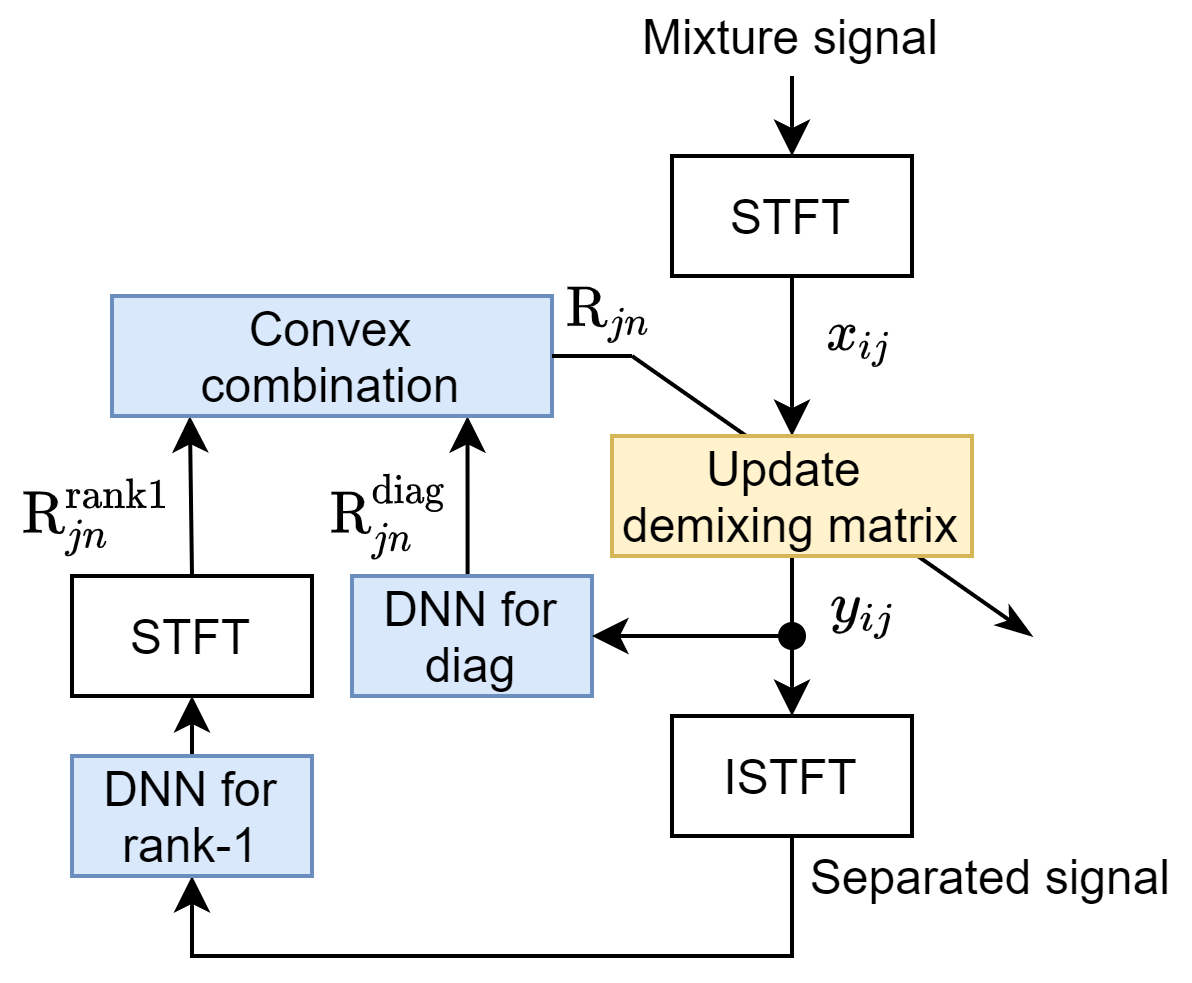}
    \vspace{-3mm}
    \caption{Schematic diagram of the proposed IDLTA.}
    \label{fig:IDLTA}
\end{figure}
We develop an algorithm for IDLTA that updates $\vb{R}_{jn}$ and $\vb{W}_i$ alternately.
Figure \ref{fig:IDLTA} shows a schematic diagram of IDLTA. The algorithm procedure is as follows:
\begin{enumerate}
    \item Initialize $\vb{W}_1,\ldots, \vb{W}_I$ as the identity matrix.
    \item Iterate the following steps until convergence:
    \begin{enumerate}
        \item Compute the separated signals as $\vb*{y}_{ij}=\vb{W}_i\vb*{x}_{ij}$.
        \item Infer FCMs as described in Section \ref{sec:model_prop}.
        \item Update $\Bqty{\vb*{w}_{in}}_{i,n}$ as described in Section \ref{sec:vcd}.
    \end{enumerate}
\end{enumerate}

\subsection{Optimization of demixing matrix $\vb{W}_i$} \label{sec:vcd}
We explain that the problem of minimizing $\mathcal{L}$ with respect to $\vb{W}_i$ when $\vb{R}_{jn}$ is kept fixed can be solved by an algorithm called vectorwise coordinate descent (VCD)\cite{Makishima:SigPro2021}, which has been developed for the conventional IPSDTA\cite{Kondo:ICASSP2020}.
VCD is a block coordinate descent method that cyclically update $\vb*{w}_{in}$ for each $i = 1,\ldots, I$ and $n = 1,\ldots, N$ one by one.
When we regard $\vb*{w}_{i'n}~(i'\neq i)$ as a constant, the objective function $\mathcal{L}$ with respect to $\vb*{w}_{in}$ is expressed as
\begin{align}
    \frac{1}{J}\mathcal{L} \ceq &\sum_{n=1}^N\pqty{\vb*{w}_{in}^\hermite\vb{Q}_{in}\vb*{w}_{in} + \vb*{w}_{in}^\hermite\vb*{\gamma}_{in} + \vb*{\gamma}_{in}^\hermite\vb*{w}_{in}} \notag\\
    &-\log \abs{\det\vb{W}_i}^2, \label{eq:cost_vcd} \\
    \vb{Q}_{in}
    =&~ \frac{1}{J}\sum_{j=1}^J\bqty{\vb{R}_{jn}^{-1}}_{ii}\vb*{x}_{ij}\vb*{x}_{ij}^\hermite , \label{eq:vcd_Q}\\
    \vb*{\gamma}_{in} 
    =&~ \frac{1}{J}\sum_{j=1}^J\pqty{\sum_{i'\in\{1,\ldots,I\}\backslash\{i\}}\bqty{\vb{R}_{jn}^{-1}}_{i'i}\vb*{x}_{i'j}^\hermite\vb*{w}_{i'n}}\vb*{x}_{ij} .\label{eq:vcd_gamma}
\end{align}
Here, $\bqty{\cdot}_{i'i}$ denotes the $\pqty{i', i}$th element of the matrix. 
As shown in \cite{Kondo:ICASSP2020}, the problem of minimizing $\mathcal{L}$ with respect to $\vb*{w}_{in}$ when all the other variables are kept fixed can be globally solved as
\begin{align}
    &\vb*{\zeta}_{in} \leftarrow \pqty{\vb{W}_i\vb{Q}_{in}}^{-1} \vb*{e}_n, \label{eq:vcd_s}\\
    &\widehat{\vb*{\zeta}}_{in} \leftarrow \vb{Q}_{in}^{-1}\vb*{\gamma}_{in}, \\
    &\eta_{in} , \leftarrow \vb*{\zeta}_{in}^\hermite\vb{Q}_{in}\vb*{\zeta}_{in}, \\
    &\widehat{\eta}_{in} \leftarrow \vb*{\zeta}_{in}^\hermite\vb{Q}_{in}\widehat{\vb*{\zeta}}_{in}, \\
    &\vb*{w}_{in} \notag \\
    &\leftarrow\left\{\begin{array}{ll}
        \frac{\vb*{\zeta}_{in}}{\sqrt{\eta_{in}}} - \widehat{\vb*{\zeta}}_{in} & (\widehat{\eta}_{in}=0) \vspace{2mm} \\
        \frac{\widehat{\eta}_{in}}{2\eta_{in}}\pqty{1 - \sqrt{1 + \frac{4\eta_{in}}{\abs{\widehat{\eta}_{in}}^2}}}\vb*{\zeta}_{in} - \widehat{\vb*{\zeta}}_{in} & (\text{otherwise}),
    \end{array}\right. \label{eq:vcd_t}
\end{align}
where $\vb*{e}_n$ is the unit vector whose $n$th element is 1 and other elements are 0.

\subsection{Acceleration of VCD}\label{sec:accel_vcd}
We here newly develop an algorithm that accelerates VCD presented in Section \ref{sec:vcd} by exploiting that $\vb{R}_{jn}$ is the sum of a diagonal matrix and a rank-1 matrix. Using the Sherman--Morrison formula \cite{Sherman:AMS1950}, we compute the $\vb{R}_{jn}^{-1}$ required in \eqref{eq:vcd_Q} and \eqref{eq:vcd_gamma} as
\begin{align}
    \vb{R}_{jn}^{-1} 
    &= \frac{1}{1-\alpha}\Bqty{\pqty{\Rdiag_{jn}}^{-1} - \alpha \widehat{\vb*{z}}_{jn}\widehat{\vb*{z}}_{jn}^\hermite}, \label{eq:R_inv}\\
    \widehat{\vb*{z}}_{jn}
    &= \xi_{jn}\cdot\pqty{\Rdiag_{jn}}^{-1}\va*{z}_{jn},\\
    \xi_{jn}
    &= \pqty{1-\alpha + \alpha\sum_{i=1}^I\frac{\abs{z_{ijn}}^2}{d_{ijn}^2}}^{-\frac{1}{2}}. \label{eq:accel_vcd_xi}
\end{align}
Here, $\Rdiag_{jn}$ and $\va*{z}_{jn}$ are defined by \eqref{eq:def_D} and \eqref{eq:def_v}, respectively. 
Note that $\vb{R}_{jn}^{-1}$ can be calculated by the element-wise inverse.
With this efficient formula, we can greatly reduce the computational cost of the original VCD in the proposed IDLTA.

\section{Experiments}
\begin{figure}
    \centering
    \includegraphics[width=1.0\linewidth]{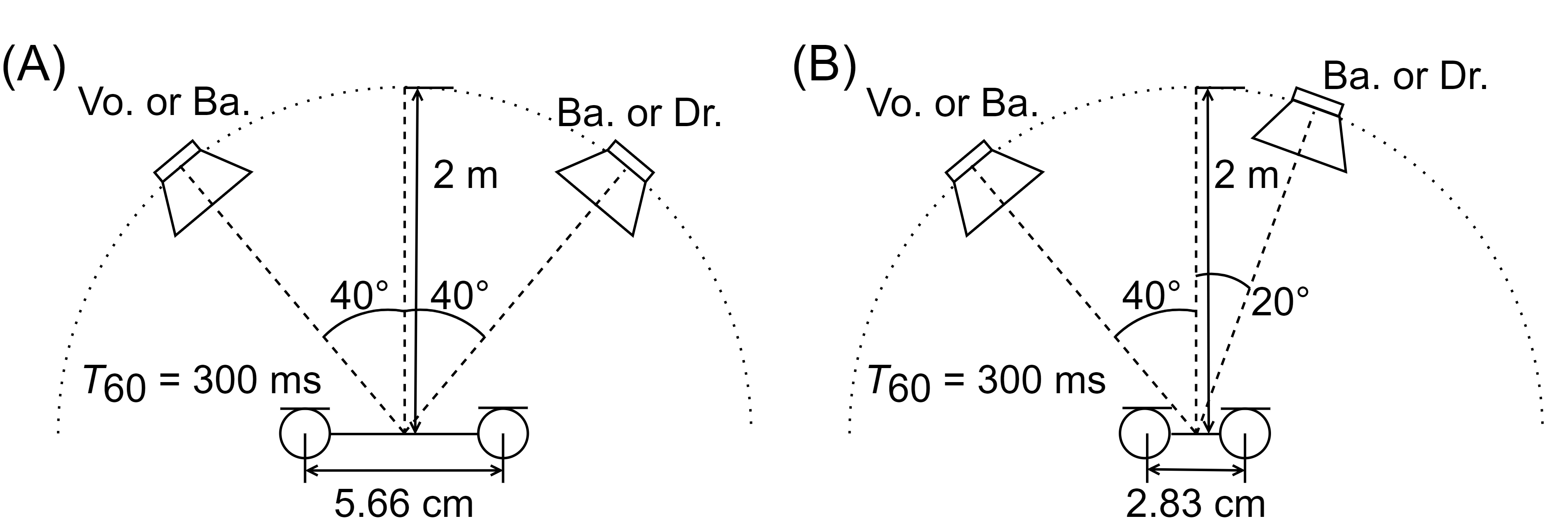}
    \vspace{-5mm}
    \caption{Spatial arrangements of sources and microphones.}
    \label{fig:rwcp}
\end{figure}

\begin{table*}[t]
\centering
\caption{Average SDR improvement of various methods [$\mathrm{dB}$]}
\label{tab:sdr_methods}
\begin{tabular}{c|cccccccccc}
\hline\hline
 & MRDLA & \multirow{2}{*}{FSCM+DNN} & \multirow{2}{*}{IDLMA} & \multicolumn{7}{c}{Proposed IDLTA} \\
 & (monaural) & & & $\alpha=0.01$ & $\alpha=0.1$ & $\alpha=0.3$ & $\alpha=0.5$ & $\alpha=0.7$ & $\alpha=0.9$ & $\alpha=0.99$ \\ \hline
Ba./Vo. & 9.97 & 12.61 & 13.65 & 14.24 & 14.56 & 15.09 & {\bf 15.28} & 15.21 & 14.66 & 10.52 \\
Vo./Dr. & 8.32 & 7.40 & 11.21 & 13.72 & 14.04 & 14.83 & {\bf 15.04} & 14.79 & 13.40 & 9.65  \\
Dr./Ba. & 5.42 & 4.39 & 5.21 & 7.08 & 8.25 & {\bf 9.16} & 9.11 & 8.44 & 6.73 & 3.94 \\\hline
average & 7.91 & 8.13 & 10.02 & 11.68 & 12.28 & 13.03 & {\bf 13.15} & 12.81 & 11.60 & 8.04 \\
\hline\hline
\end{tabular}
\end{table*}

\subsection{Conditions}\label{sec:exp_condition}
We carried out an experiment of separating music signals in the case of $N = M = 2$, and compared the following four methods:
\begin{itemize}
    \item MRDLA \cite{Nakamura:ICASSP2020,Kozuka:INTERNOISE2020}: a monaural separation DNN. (Unlike \cite{Nakamura:ICASSP2020}, the input and output of MRDLA are monaural and the sampling frequency is 8 kHz.)
    \item IDLMA \cite{Makishima:IEEE_ACM_J_ASLP2019,Makishima:SigPro2021}: identical to IDLTA with $\alpha=0$.
    \item Proposed IDLTA.
    \item FSCM+DNN \cite{Nugraha:IEEE_ACM_J_ASLP2016}: the well-known supervised method that estimates the full-rank spatial covariance matrix (FSCM), which represents the mixing system, using the source spectra inferred by $\dnndiag$.
\end{itemize}
We implemented $\dnndiag$ in IDLMA and IDLTA in the same manner as in \cite{Makishima:IEEE_ACM_J_ASLP2019}.
We used MRDLA as $\dnntd$.

We used the MUSDB18 dataset \cite{musdb18} for DNN training and performance evaluation.
100 songs were used to train DNNs, and an other 25 songs were used to evaluate the performance.
All signals were down-sampled to 8 kHz.
In the evaluation, the length of the observation signal was 30 s.
To simulate reverberant signals, the impulse responses of E2A ($\mathrm{T}_{60}=300$ ms) obtained from RWCP Sound Scene Database in Real Acoustical Environments \cite{Nakamura:LREC2000} were convoluted with each source signal.
We prepared three types of mixture: bass and vocals (Ba./Vo.); vocals and drums (Vo./Dr.); and drums and bass (Dr./Ba.).
The spatial arrangements of sources and microphones for synthesized observation signals are shown in Fig. \ref{fig:rwcp}.

In MRDLA, the reference channel signal of the multichannel observation was used as the monaural input.
In IDLMA, IDLTA, and FSCM+DNN, the number of iterations to update the spatial parameters (demixing matrices and FSCMs) was set to 100, and FCMs were updated after every 10 iterations of updating the spatial parameters.
In IDLMA and IDLTA, the demixing matrix $\vb{W}_i$ was initialized as the identity matrix.
The STFT was performed using a 512-ms-long Hamming window and a 256-ms-long shift.
In all methods, the performance was evaluated using the average signal-to-distortion ratio (SDR)\cite{Vincent:IEEE_ACM_J_ASLP2006} improvement over all songs and spatial arrangements.

\subsection{Results}

Table \ref{tab:sdr_methods} shows the comparison of the SDR improvement among MRDLA, IDLMA, and IDLTA.
In IDLTA, FCMs are inferred using DNNs as described in \ref{sec:exp_condition}.
Except for $\alpha=0.99$, IDLTA outperformed the separation performance of MRLDA and IDLMA.
In particular, the highest SDR improvement was observed at $\alpha=0.5$ for Ba./Vo. and Vo./Dr. and at $\alpha=0.3$ for Dr./Ba.
This result confirms the effectiveness of IDLTA considering the inter-frequency correlation.
The decrease in separation performance for $\alpha=0.99$ can be attributed to the fact that $\vb{R}_{jn}$ is almost $\va*{z}_{jn}\va*{z}_{jn}^\hermite$ and the separated signal is estimated only in the direction close to $\va*{z}_{jn}$, as described in Section \ref{sec:model_opt}. The performance of $\alpha=0.99$ is almost the same as that of MRDLA, which confirms the validity of this discussion.

\section{Conclusion}
We proposed a new supervised audio source separation method called IDLTA, which is an extension of the conventional IDLMA to account for the inter-frequency correlation.
It is difficult to directly infer FCMs using DNNs.
Therefore, we also proposed a new FCM of source spectra, which is modeled as a convex combination of two simple FCMs, and these two FCMs are inferred using DNNs.
By a numerical experiment of separating music signals, we confirmed that 
IDLTA improves the separation performance of IDLMA.

\bibliographystyle{IEEEtran}
\bibliography{reference.bib}

\end{document}